\documentclass[%
reprint,
superscriptaddress,
showpacs,
nofootinbib,
 amsmath,amssymb,
 aps,   
prl,
]{revtex4-2}

\usepackage{amsmath}
\usepackage{amsfonts} 
\usepackage{mathtools} 
\usepackage{amssymb} 
\usepackage{subfig} 
\usepackage{graphicx}
\usepackage{xcolor}
\usepackage{mathrsfs}
\usepackage[breaklinks,colorlinks,urlcolor=blue,citecolor=blue,linkcolor=magenta]{hyperref}
\usepackage{multirow}
\usepackage{orcidlink}

\newcommand{\refEq}[1]{Eq.\,\eqref{#1}} 
\newcommand{\refEqs}[1]{Eqs.\,\eqref{#1}} 
\newcommand{\re}[1]{\text{Re}\left(#1\right)}
\newcommand{\im}[1]{\text{Im}\left(#1\right)}

\newcommand{\SD}[1]{\Phi_{#1}^{\phantom{\dagger}}}
\newcommand{\SDd}[1]{\Phi_{#1}^\dagger}
\newcommand{\SDc}[1]{\Phi_{#1}^\ast}

\newcommand{\bilH}[1]{\mathcal H_{#1}}
\newcommand{\bilA}[1]{\mathcal A_{#1}}

\newcommand{\qq}[1]{\mu^2_{#1}}
\newcommand{\QQ}[1]{\lambda_{#1}}
\newcommand{\QQA}[1]{\lambda^{\mathcal{A}}_{#1}}

\newcommand{\nfR}[1]{\mathrm{R}_{#1}}
\newcommand{\nfI}[1]{\mathrm{I}_{#1}}
\newcommand{\chfPM}[1]{\mathrm{C}^\pm_{#1}}
\newcommand{\chfP}[1]{\mathrm{C}^+_{#1}}
\newcommand{\chfM}[1]{\mathrm{C}^-_{#1}}
\newcommand{\chfvec}{{\vec{\mathbf{C}}}}
\newcommand{\nfvec}{{\vec{\mathbf{N}}}}

\newcommand{\vev}[1]{v_{#1}}
\newcommand{\vevPh}[1]{\theta_{#1}}

\newcommand{\nMM}{M_0^2}

\newcommand{\chMM}{M_{\pm}^2}

\newcommand{\potV}{\mathcal{V}}

\graphicspath{{Figs/}}

\begin{document}

\title{Light states in real multi-Higgs models with spontaneous CP violation}

\author{Carlos Miró}\email{Carlos.Miro@uv.es}
\author{Miguel Nebot}\email{Miguel.Nebot@uv.es}
\author{Daniel Queiroz}\email{Daniel.Queiroz@uv.es}
\affiliation{%
Departament de Física Teòrica \& Instituto de Física Corpuscular (IFIC),\\ Universitat de València -- CSIC, E-46100 Valencia, Spain
}


\begin{abstract}
In models with extended scalar sectors consisting of multiple Higgs doublets that trigger spontaneous electroweak symmetry breaking, 
it might be expected that the abundance of dimensionful quadratic couplings in the scalar potential could allow for a regime where, apart from the would-be Goldstone bosons and a neutral Higgs-like state, all new scalars have masses much larger than the electroweak scale. In the case of models where CP invariance holds at the lagrangian level but is broken by the vacuum, we show that such a reasonable expectation does not hold. When perturbativity requirements are placed on the dimensionless quartic couplings, the spectrum of the new scalars includes one charged and two additional neutral states whose masses cannot be much larger than the electroweak scale.
\end{abstract}

\maketitle

\section{Introduction\label{SEC:Intro}}
The quest for a better understanding of open fundamental questions often involves new particles. Knowing their masses is undoubtedly a priority. As the decades-long history of the Higgs boson shows, short of the experimental grail of discovery \cite{ATLAS:2012yve,CMS:2012qbp} or indirect hints \cite{Baak:2011ze}, theoretical arguments can be put forward to bring light on this respect \cite{Weinberg:1976pe,Politzer:1978ic,Cabibbo:1979ay,Dashen:1983ts,Callaway:1983zd}. An example of them is the use of perturbativity requirements in the classical works of B. Lee, Quigg and Thacker \cite{Lee:1977yc,Lee:1977eg} (see also \cite{Dicus:1973gbw} for earlier work) to obtain bounds on the Higgs mass. An important implication of perturbativity requirements on extended scalar sectors featuring spontaneous symmetry breaking, is the existence of a Higgs-like state whose mass cannot be (much) larger than the electroweak scale --given by the vacuum expectation value (vev) $\vev{}\simeq 246$ GeV--, as discussed in \cite{Langacker:1984dma,Weldon:1984th}. Two Higgs doublets models (2HDMs) were proposed by T.D. Lee \cite{Lee:1973iz,Lee:1974jb}, with the appealing possibility of a spontaneous origin of CP violation, that is, having a CP-invariant Lagrangian together with a CP-non-invariant vacuum; for reasons to become clear later, we refer to them as \emph{real} 2HDMs with spontaneous CP violation (SCPV). Besides the Higgs-like state, perturbativity-based bounds on the masses of the additional scalars have attracted attention within 2HDMs endowed with some symmetry \cite{Huffel:1980sk,Casalbuoni:1987cz,Maalampi:1991fb,Kanemura:1993hm,Ginzburg:2005dt,Horejsi:2005da,Kanemura:2015ska} --not of the CP type, typically a $\mathbb{Z}_2$ symmetry--. Complementarily, focusing on multi-Higgs models \cite{Ivanov:2017dad}, the analysis of scenarios with new \emph{heavy} scalars, i.e. scalars with masses $\gg\vev{}$, was addressed generically in \cite{Haber:1989xc}, and recently in \cite{Faro:2020qyp,Carrolo:2021euy} in connection with symmetries (again not of the CP type). Concerning real 2HDMs with SCPV, it was realized in \cite{Nebot:2018nqn}, and analyzed in detail in \cite{Nebot:2019qvr}, that those perturbativity-based bounds on the masses apply to \emph{all} the new scalars in the model, one charged and two neutral ones --in addition, of course, to the neutral Higgs-like state--. For some phenomenological implications see \cite{Nierste:2019fbx}; for earlier work on the connection between the vacuum CP properties and the mass spectrum in extended scalar sectors see \cite{Barenboim:2001vu}.\\ 
The rationale is rather simple. With spontaneous symmetry breaking, there are two types of mass terms in 2HDMs, either dimensionful quadratic couplings or dimensionless quartic couplings $\times$ (vevs)$^2$. If the quartic couplings are limited by perturbativity requirements, masses (much) larger than $\vev{}$ can only arise through correspondingly large quadratic couplings. The crucial particularity of the real 2HDM with SCPV is the fact that there are only 3 quadratic couplings, which is also the number of stationarity conditions imposed on the scalar potential in order that the vacuum is an extremum. These stationarity conditions can then be used to trade all 3 quadratic couplings for quartic couplings $\times$ (vevs)$^2$. As a consequence, all mass terms are bounded through perturbativity requirements on the quartic couplings. This outcome is indeed peculiar: rather than the generic expectation of having at least one light scalar --the Higgs-like one--, perturbativity requirements bound \emph{the whole spectrum}. This opens a very relevant question: if instead of a real 2HDM with SCPV, one considers a real $n$HDM with SCPV, is there something similar at work?\\ 
The prospects on that respect are, a priori, discouraging: the number of quadratic couplings scales with $n^2$ while the number of stationarity conditions only scales with $n$. The case $n=2$ is indeed very peculiar: for $n>2$, free quadratic couplings are necessarily present in the scalar potential. Does this imply that all the masses of the new scalars can be much larger than the electroweak scale?

The central result of this work is that the answer to this question is in the negative: \emph{for all $n$, the spectrum necessarily includes one charged and two neutral scalars (in addition to the neutral Higgs-like) that must be light, that is, whose masses cannot be much larger than the electroweak scale when perturbativity requirements are imposed on the quartic couplings}.  

The manuscript is organized as follows. We introduce first the real $n$HDM with SCPV, proceed then with a numerical exercise that will illustrate the main points in a kind of phenomenological exploration, and finally address analytically the central question of this work.

\section{Real $n$HDM with SCPV\label{SEC:RealnHDMSCPV}}
For $n$ Higgs doublets $\SD{a}$, $a=1,\ldots,n$, the most general scalar potential invariant under the CP transformation $\SD{a}\mapsto\SDc{a}$ has the following form:
\begin{equation}\label{eq:RealnHDM:Pot:01}
 \potV(\SD{1},\ldots,\SD{n})=\potV_2(\SD{1},\ldots,\SD{n})+\potV_4(\SD{1},\ldots,\SD{n}),
\end{equation}
with
\begin{equation}\label{eq:RealnHDM:Pot:V2:01}
 \potV_2(\SD{1},\ldots,\SD{n})=\sum_{a=1}^n\qq{a}\SDd{a}\SD{a}+\sum_{a=1}^{n-1}\sum_{b=a+1}^n\qq{ab}\bilH{ab},
\end{equation}
\begin{equation}\label{eq:RealnHDM:Pot:V4:01}
 \begin{aligned}
 &\potV_4(\SD{1},\ldots,\SD{n})=\\
 &\sum_{a=1}^n\QQ{a}(\SDd{a}\SD{a})^2+\sum_{a=1}^{n-1}\sum_{b=a+1}^{n}\QQ{a,b}(\SDd{a}\SD{a})(\SDd{b}\SD{b})\\
 &+\sum_{a=1}^n\sum_{b=1}^{n-1}\sum_{c=b+1}^n\QQ{a,bc}(\SDd{a}\SD{a})\bilH{bc}\\
 &+\left.\sum_{a=1}^{n-1}\sum_{b=a+1}^n\sum_{c=1}^{n-1}\sum_{d=c+1}^n\right|_{(a,b)\leq(c,d)}\hspace{-8ex}\left(\QQ{ab,cd}\bilH{ab}\bilH{cd}+\QQA{ab,cd}\bilA{ab}\bilA{cd}\right).\end{aligned}
\end{equation}
All quadratic $\qq{a}$, $\qq{ab}$ in $\potV_2$, and quartic $\QQ{a}$, $\QQ{a,b}$, $\QQ{a,bc}$, $\QQ{ab,cd}$, $\QQA{ab,cd}$ in $\potV_4$, parameters are real (hence \emph{real} $n$HDM). We use hermitian and antihermitian  bilinears ($a<b$)
\begin{equation}
 \bilH{ab}\equiv\frac{1}{2}(\SDd{a}\SD{b}+\SDd{b}\SD{a}),\ \bilA{ab}\equiv\frac{1}{2}(\SDd{a}\SD{b}-\SDd{b}\SD{a}).
\end{equation}
Under the CP transformation $\SD{a}\mapsto\SDc{a}$, $\bilH{ab}\mapsto\bilH{ab}$ and $\bilA{ab}\mapsto -\bilA{ab}$: invariance of the potential is immediately transparent. Notice that in the last terms in \refEq{eq:RealnHDM:Pot:V4:01} the ordering $(a,b)\leq(c,d)$ stands for $a<c$ and, in case $a=c$, $b<d$, and is introduced to avoid repeated terms in the sums.\\
Assuming an appropriate electroweak symmetry breaking vacuum, expansion of fields reads
\begin{equation}\label{eq:Fieldvev:01}
 \SD{a}=\frac{e^{i\vevPh{a}}}{\sqrt 2}\begin{pmatrix}\sqrt{2}\chfP{a}\\ \vev{a}+\nfR{a}+i\,\nfI{a}\end{pmatrix},\quad \langle\SD{a}\rangle=\frac{\vev{a}e^{i\vevPh{a}}}{\sqrt 2}\begin{pmatrix}0\\ 1\end{pmatrix},
\end{equation}
where the vevs $\langle\SD{a}\rangle$ are parameterized by $\vev{a}\in\mathbb{R}^+$, $\vev{1}^2+\ldots+\vev{n}^2=\vev{}^2\simeq 246^2$ GeV$^2$, and $\vevPh{a}\in[0;2\pi[$, moduli and phases respectively. Individual phases have no physical meaning: all CP violation arising from the vacuum is encoded in the phase differences $\vevPh{a}-\vevPh{b}$. To obtain the stationarity conditions, one first computes
\begin{equation}
 V(\vev{1},\ldots,\vev{n},\vevPh{1},\ldots,\vevPh{n})=\potV(\langle\SD{1}\rangle,\ldots,\langle\SD{n}\rangle),
\end{equation}
and then sets derivatives with respect to the vev parameters to zero,
\begin{equation}\label{eq:Stationarity:01}
 \partial_{\vev{1}}V=\ldots=\partial_{\vev{n}}V=0,\ 
 \partial_{\vevPh{1}}V=\ldots=\partial_{\vevPh{n}}V=0,
\end{equation}
with $\partial_xV\equiv\frac{\partial V}{\partial x}$. The lack of physical meaning of individual phases translates into $\partial_{\vevPh{1}}V+\ldots+\partial_{\vevPh{n}}V=0$ \emph{by construction}, irrespective of imposing that each term is zero: \refEqs{eq:Stationarity:01} give $2n-1$ (independent) stationarity conditions. With \refEqs{eq:RealnHDM:Pot:V2:01}-\eqref{eq:RealnHDM:Pot:V4:01}, the derivatives in \refEqs{eq:Stationarity:01} read
\begin{equation}\label{eq:Stationarity:v:01}
\begin{aligned}
 &\partial_{\vev{1}}{V}=\qq{1}\vev{1}+\frac{1}{2}\sum_{b=2}^n\qq{1b}c_{1b}\vev{b}+[\lambda's],\\
 &\partial_{\vev{j}}{V}=\qq{j}\vev{j}+\frac{1}{2}\sum_{a=1}^{j-1}\qq{aj}c_{aj}\vev{a}+\frac{1}{2}\sum_{b=j+1}^n\qq{jb}c_{jb}\vev{b}+[\lambda's],\\
 &\partial_{\vev{n}}{V}=\qq{n}\vev{n}+\frac{1}{2}\sum_{a=1}^{n-1}\qq{an}c_{an}\vev{a}+[\lambda's],
\end{aligned}
\end{equation}
\begin{equation}\label{eq:Stationarity:th:01}
\begin{aligned}
 &\partial_{\vevPh{1}}{V}=-\frac{1}{2}\sum_{b=2}^n\qq{1b} s_{1b}\vev{1}\vev{b}+[\lambda's],\\
 &\partial_{\vevPh{j}}{V}=\frac{1}{2}\sum_{a=1}^{j-1}\qq{aj}s_{aj}\vev{a}\vev{j}-\frac{1}{2}\sum_{b=j+1}^n\qq{jb}s_{jb}\vev{j}\vev{b}+[\lambda's],\\
 &\partial_{\vevPh{n}}{V}=\frac{1}{2}\sum_{a=1}^{n-1}\qq{an}s_{an}\vev{a}\vev{n}+[\lambda's],
\end{aligned}
\end{equation}
where the shorthand notation $\vevPh{ab}=\vevPh{a}-\vevPh{b}$, $c_{ab}=\cos\vevPh{ab}$, $s_{ab}=\sin\vevPh{ab}$, is used, and $[\lambda's]$ stand for terms involving only quartic couplings, not displayed for conciseness.\\ 
The mass terms within $\potV\supset-\mathscr L_{\mathrm{Mass}}$ are
\begin{equation}\label{eq:V:massterms:01}
 -\mathscr L_{\mathrm{Mass}}=\chfvec^\dagger\,\chMM\,\chfvec+\frac{1}{2}\nfvec^T\nMM\nfvec,
\end{equation}
with $\chfvec^\dagger=(\chfP{1},\ldots,\chfP{n})$, $\nfvec^T=(\nfR{1},\ldots,\nfR{n},\nfI{a},\ldots,\nfI{n})$. $\chMM$ is
the $n\times n$ charged mass matrix and $\nMM$ the $2n\times 2n$ neutral mass matrix, whose elements are
\begin{equation}
\begin{aligned}
&(\chMM)_{a,b}=\left[\frac{\partial^2\potV}{\partial\chfP{a} \partial\chfM{b}}\right],\\
&(\nMM)_{a,b}=\left[\frac{\partial^2\potV}{\partial\nfR{a} \partial\nfR{b}}\right],\ 
(\nMM)_{n+a,n+b}=\left[\frac{\partial^2\potV}{\partial\nfI{a} \partial\nfI{b}}\right],\\
&(\nMM)_{a,n+b}=(\nMM)_{n+b,a}=\left[\frac{\partial^2\potV}{\partial\nfR{a} \partial\nfI{b}}\right],
\end{aligned}
\end{equation}
where $[\ ]$ above are evaluated at $\chfPM{a}$, $\nfR{a}$, $\nfI{a}\to 0$. 
Focusing on the number of quadratic couplings, $\potV_2$ in \refEq{eq:RealnHDM:Pot:V2:01} has $n(n+1)/2$ of them ($n$ $\qq{a}$ and $n(n-1)/2$ $\qq{ab}$ with $a<b$) to be compared with the $2n-1$ (independent) stationarity conditions. For $n=2$, as already mentioned, both numbers match and the stationarity conditions can be used to trade all quadratic couplings for quartic couplings $\times$ vevs. For $n>2$, one is quickly driven into an overabundance of quadratic couplings with respect to stationarity conditions, as illustrated in Table \ref{tab:count}, which shows that for $n=3$ there is already one free quadratic coupling, for $n=5$ there are more quadratic couplings than charged scalars, and for $n=7$ there are more free quadratic couplings than neutral scalars. One could expect that, beyond $n=2$, the existing free quadratic couplings can drive arbitrarily large \emph{new} scalar masses, \emph{new} meaning neither the would-be Goldstone bosons (wbG hereafter) nor the light Higgs-like state. As we first illustrate numerically and then address analytically, that reasonable expectation is surprisingly misled.
\begin{table}[!htb]
\begin{center}
\begin{tabular}{|c||c|c|c|c|c|c|}\hline
 $n$ & \,2\, & \,3\, & 4 & 5 & 6 & 7\\\hline
 $n(n+1)/2$ & 3 & 6 & 10 & 15 & 21 & 28\\\hline
 $2n-1$ & 3 & 5 & 7 & 9 & 11 & 13\\\hline
\end{tabular}
\end{center}
\vspace{-1ex}
\caption{\label{tab:count}Number of quadratic couplings, $n(n+1)/2$, and of stationarity conditions, $2n-1$, as a function of $n$. Not counting the would-be Goldstone bosons, there are $n-1$ charged and $2n-1$ neutral scalars.}
\end{table}
%

\section{Numerical exercise\label{SEC:Num}}
\begin{figure*}[ht]
	\centering
    \subfloat[Fixed $k_\lambda=4$.\label{fig:num:Ch:lambda}]{\includegraphics[width=0.4\textwidth]{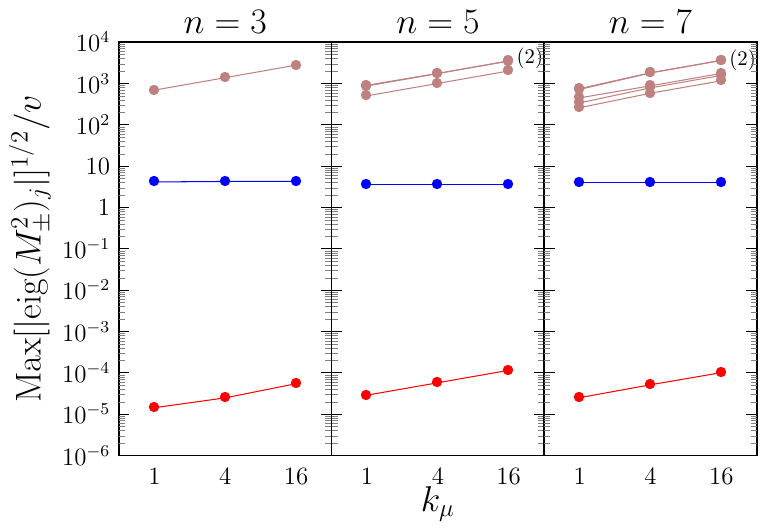}}\qquad 
    \subfloat[Fixed $k_\mu=4$.\label{fig:num:Ch:mu}]{\includegraphics[width=0.4\textwidth]{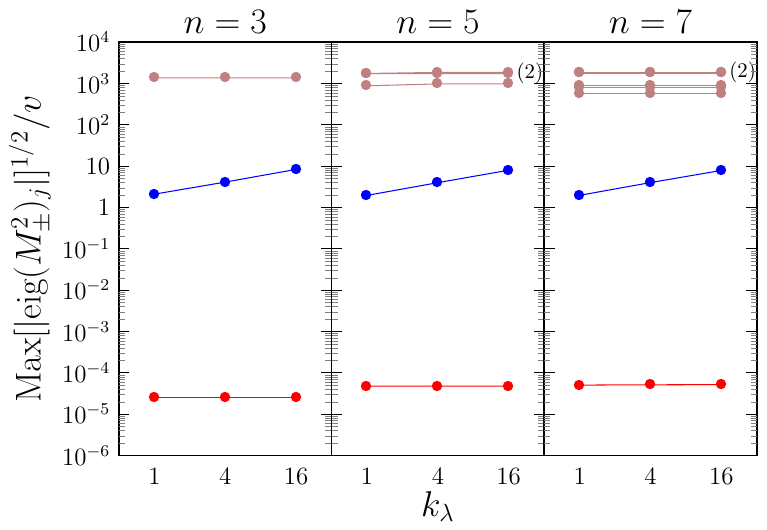}}
    \caption{Maximal eigenvalues of $\chMM$ for 3, 5 and 7HDMs. For the heavier states (brown dots and lines), ``(2)'' inserts indicate near-degeneracies (number of states represented). \label{fig:num:Ch}} 
\end{figure*}

\begin{figure*}[ht]
	\centering
    \subfloat[Fixed $k_\lambda=4$.\label{fig:num:N:lambda}]{\includegraphics[width=0.4\textwidth]{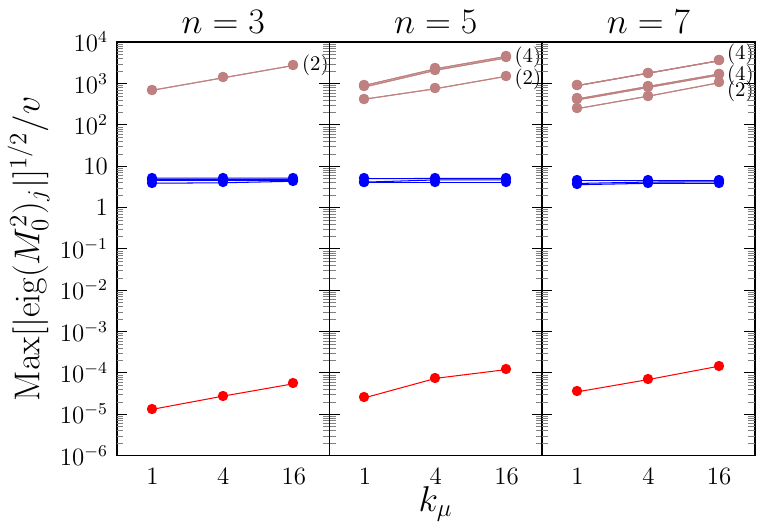}}\qquad
    \subfloat[Fixed $k_\mu=4$.\label{fig:num:N:mu}]{\includegraphics[width=0.4\textwidth]{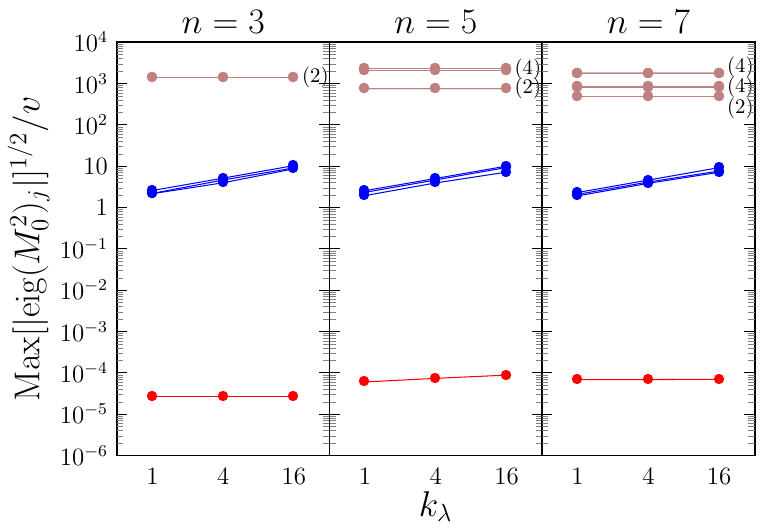}}
    \caption{Maximal eigenvalues of $\nMM$ for 3, 5 and 7HDMs. For the heavier states (brown dots and lines), ``(2)'' and ``(4)'' inserts indicate near-degeneracies (number of states represented).\label{fig:num:N}} 
\end{figure*}

For a given number $n$ of Higgs doublets (we will consider the cases $n=3,5,7$), we perform a simple numerical exercise as follows.\\
$\bullet$ We use the stationarity conditions in \refEqs{eq:Stationarity:th:01} to trade all $n-1$ $\qq{1j}$ for other quadratic $\qq{ij}$  ($i=2,\ldots,n-1$, $j=i+1,\ldots,n$) and quartic parameters.\\
$\bullet$ We use them again, together with \refEqs{eq:Stationarity:v:01} to trade all $n$ $\qq{j}$ for other quadratic $\qq{ij}$ ($i=2$ to $n-1$, $j=i+1$ to $n$) and quartic parameters.\\
$\bullet$ We are left with $n(n+1)/2-(2n-1)=(n-1)(n-2)/2$ free $\qq{ij}$, $i=2$ to $n-1$, $j=i+1$ to $n$. We generate random values for them in the range $[-k_\mu;k_\mu]\times 10^{10}$ GeV$^2$, for 3 separate values $k_\mu=1, 4, 16$.\\
$\bullet$ We generate random values of all quartic couplings --$\QQ{}$'s in \refEq{eq:RealnHDM:Pot:V4:01}-- in the range $[-k_\lambda;k_\lambda]$, separately for $k_\lambda=1,4,16$ (the purpose of considering different $k_\mu$ and $k_\lambda$ will become clear when discussing the results).\\
$\bullet$ We generate random values of the moduli of the vevs $\vev{j}$ in a range $[0;1]$ GeV and then rescale them collectively to fix $\vev{1}^2+\ldots+\vev{n}^2=\vev{}^2\simeq 246^2$ GeV$^2$.\\
$\bullet$ We generate random real values of the phases of the vevs $\vevPh{j}$ in a range $[0;2\pi[$.\\
$\bullet$ With these numerical values, we compute $\qq{j}$, $j=1$ to $n$, and $\qq{1j}$, $j=2$ to $n$, and check if they are all in the range $[-k_\mu;k_\mu]\times 10^{10}$ GeV$^2$: if they are, we continue, otherwise the procedure up to this point is repeated.\\
$\bullet$ We compute the mass matrices $\chMM$ and $\nMM$.\\
$\bullet$ We compute their eigenvalues and sort them according to their absolute values. Notice that we do not require \emph{(i)} the eigenvalues to be non-negative (that would amount to requiring that the vacuum is a local minimum), \emph{(ii)} the quartic couplings to verify boundedness from below of the potential. There is no need to sound the alarm on these respects considering the scope and results of this numerical exercise, as we comment below.\\
$\bullet$ The whole procedure is repeated, keeping the largest value obtained for each of the sorted absolute values of the eigenvalues of $\chMM$ and $\nMM$. The number of repetitions ranges from $\mathcal O(10^6)$ for $n=3$ to $\mathcal O(10^3)$ for $n=7$, attending to the required computing time, which increases non-linearly with $n$. These largest values are then represented (normalized to the electroweak vev $\vev{}$). The results are shown in Fig. \ref{fig:num:Ch} for the charged scalars, and Fig. \ref{fig:num:N} for the neutral ones. Within each figure, the left subfigure corresponds to a fixed value $k_\mu=4$, and the 3 different values of $k_\lambda$, while the opposite is presented in the right subfigure, that is, a fixed value $k_\lambda=4$ and the 3 different values of $k_\mu$.


\noindent The important points to notice are the following.
\begin{enumerate}
 \item The wbG, represented in red, are readily identified (rather than the analytic eigenvalue zero, they yield values as small as expected considering the finite numerical precision).
 \item There is one charged light state (represented in blue in Fig.~\ref{fig:num:Ch}), whose mass cannot exceed $\mathcal O(\vev{})$, is sensitive to $k_\lambda$, and insensitive to $k_\mu$: notice the slope of the blue line in Fig.~\ref{fig:num:Ch:mu}, to be compared with the horizontal one in Fig.~\ref{fig:num:Ch:lambda}. Similarly, there are 3 neutral light states (represented in blue in Fig.~\ref{fig:num:N}), whose masses cannot exceed $\mathcal O(\vev{})$, which are sensitive to $k_\lambda$ and insensitive to $k_\mu$. Besides the expected light neutral Higgs-like, these states, one charged and two neutrals, are highly unexpected.
 \item There are $n-2$ charged and $2n-4$ neutral heavy states, represented in brown in Figs.~\ref{fig:num:Ch} and \ref{fig:num:N} respectively, whose masses can largely exceed $\mathcal O(\vev{})$, are insensitive to $k_\lambda$, and sensitive to $k_\mu$.
\end{enumerate}
As a final comment, it is now clear why we could dispense of the additional physical requirements of non-negative eigenvalues and bounded from below potential. The numerical results reveal the unexpected existence of necessarily-light states: that property also holds in the physically meaningful subspace of the considered parameter space. In any case, one could have performed this numerical exercise including the physical requirements but, \emph{(i)} the boundedness from below constraint on the quartic couplings for a generic number of doublets has no analytic answer in the literature, \emph{(ii)} it would have a prohibitive computing time cost (we checked that just requiring non-negative eigenvalues already has a very significant impact, with no apparent effect on the results of interest). All in all, the aim of this numerical exercise is to illustrate the unexpected appearance of states that cannot have masses above the electroweak scale, and to provide some leads into an analytic understanding of the situation, to which we now turn.

\section{Analysis\label{SEC:Analysis}}
If one were capable of carrying out analytically the obtention of the eigenvalues of the charged and neutral scalar mass matrices for an arbitrary number of doublets, one would simply read out that, unexpectedly, apart from the massless wbG and the Higgs-like scalar, there are one charged and two neutral scalars that are light, i.e. with masses not exceeding $\mathcal O(\vev{})$.
Short of that kind of capacity, we need to think our way into a better understanding, assisted by the previous numerical insights.
First, driving the new scalar masses to a regime of large values requires large quadratic terms $\gg\vev{}^2$. In that case, it might be worth considering the stationarity conditions and the mass matrices without quartic couplings \emph{at all}. The fact that in our numerical exercise the unexpected light states have eigenvalues apparently insensitive to $k_\mu$ hints to the possibility that these eigenvalues are independent of the quadratic couplings. Then, considering the whole problem without quartic couplings, this can only make sense if these unexpected states appear as eigenstates with eigenvalues equal to zero, i.e. null eigenvectors. 
That is indeed the case, and the gentler path that we will try to follow to the desired result is: \emph{(i)} write down the mass matrices in the mentioned regime in which quartic couplings are dropped, \emph{(ii)} notice a property that reduces the problem of dealing with both charged $n\times n$ and neutral $2n\times 2n$ mass matrices to just dealing with the charged mass matrix, \emph{(iii)} analyze the case of the wbG for a better understanding or inspiration.
One can then think of the complete problem including quartic couplings as a (degenerate) perturbation theory problem where the contributions to the entries of the mass matrices from the quartic couplings are the perturbation. 
The important point to anticipate is that, consequently, null eigenvectors of the quadratic-couplings-alone mass matrices cannot yield scalars with squared masses much larger than the $\QQ{}\times\vev{}^2$ perturbation.

Without quartic couplings, the scalar potential in \refEqs{eq:RealnHDM:Pot:01}-\eqref{eq:RealnHDM:Pot:V4:01} is simply $\potV\to\potV_2$.
The stationarity conditions are as in \refEqs{eq:Stationarity:v:01}-\eqref{eq:Stationarity:th:01} with $[\lambda's]\to 0$. Not using (yet) the stationarity conditions, one can read the mass terms $-\mathscr L_{\mathrm{Mass}}\subset\potV_2$: 
\begin{equation}
\begin{aligned}
& [\chMM]_{a,a}=\qq{a},\ [\chMM]_{a,b}=[\chMM]_{b,a}^\ast=\frac{1}{2}e^{i\vevPh{ab}}\qq{ab},\ a<b,\\
& \chMM=
 \begin{pmatrix}
 \qq{1} & \frac{1}{2}e^{i\vevPh{12}}\qq{12} & \cdots &  \\ 
  \frac{1}{2}e^{-i\vevPh{12}}\qq{12} & \qq{2} & \cdots & \\
  \vdots & \vdots & \ddots & \\ 
   &  & & \qq{n}
  \end{pmatrix}.
\end{aligned}
\end{equation}
Crucially, $\chMM$ and $\nMM$ are related easily:
\begin{equation}
\nMM =\begin{pmatrix} \re{\chMM} & \im{\chMM} \\ -\im{\chMM} & \re{\chMM}\end{pmatrix},
\end{equation}
with $\re{\chMM}^T=\re{\chMM}$, $\im{\chMM}^T=-\im{\chMM}$.
Now, if $\chMM$ has a null eigenvector $\vec c\in\mathbb{C}^n$, $\chMM\,\vec c=\vec 0_n$ ($\vec 0_n$ is the $n$-vector with all components $0$); expanding real and imaginary parts, one immediately obtains
\begin{equation}\label{eq:MCh:Null:01}
\begin{aligned}
 &\re{\chMM}\re{\vec c}-\im{\chMM}\im{\vec c}=\vec 0_n,\\
 &\im{\chMM}\re{\vec c}+\re{\chMM}\im{\vec c}=\vec 0_n.
\end{aligned}
\end{equation}
In matrix form, \refEqs{eq:MCh:Null:01} give
 $\nMM \vec r_1=\vec 0_{2n}$, $\nMM\vec r_2=\vec 0_{2n}$, 
\begin{equation}\label{eq:MChMN:Null:01}
\text{with}\quad \vec r_1=\begin{pmatrix} \re{\vec c}\\ -\im{\vec c}\end{pmatrix},\quad 
 \vec r_2=\begin{pmatrix} \im{\vec c}\\ \re{\vec c}\end{pmatrix}. 
\end{equation}
From \emph{one} null eigenvector $\vec c$ of $\chMM$, we obtain the \emph{two} null eigenvectors $\vec r_1$ and $\vec r_2$ of $\nMM$ in \refEq{eq:MChMN:Null:01} (notice in addition that $\vec r_1$ and $\vec r_2$ are orthogonal).
We already know one null eigenvector $\vec c_{G}$ of $\chMM$, corresponding to the charged wbG, $\vec c_{G}^{\,T}=(\vev{1},\ldots,\vev{n})$. It is instructive to observe that 
 \begin{equation}
 \chMM \vec c_{G}=
  \begin{pmatrix}
  \partial_{\vev{1}}{V}_2-\frac{i}{\vev{1}}\partial_{\vevPh{1}}{V}_2\\
  \vdots\\
  \partial_{\vev{n}}{V}_2-\frac{i}{\vev{n}}\partial_{\vevPh{n}}{V}_2
 \end{pmatrix}.
 \end{equation}
The point of this simple check is now clear: one can read $\chMM \vec c_{G}=\vec 0_n$ transparently, since all entries $\partial_{\vev{j}}{V}_2-\frac{i}{\vev{j}}\partial_{\vevPh{j}}{V}_2$ are expressed as combinations of the derivatives in the stationarity conditions in \refEqs{eq:Stationarity:v:01}-\eqref{eq:Stationarity:th:01}.
In terms of the doubling of $\vec c_{G}$ in two null eigenvectors of $\nMM$, we have $\nMM\vec r_G=\nMM\vec r_h=\vec 0_{2n}$ with
\begin{align}
 &\vec r_G^{\,T}=\begin{pmatrix}\vec 0_n,& \vev{1},& \ldots,& \vev{n}\end{pmatrix},\ \vec r_h^{\,T}=\begin{pmatrix} \vev{1},& \ldots,& \vev{n}, & \vec 0_n\end{pmatrix}.\label{eq:NullVecNGold:01}
\end{align}
The vector $\vec r_G$ corresponds to the neutral wbG and $\vec r_h$ to the Higgs-like scalar.
The key question is then: \emph{is there another null eigenvector of $\chMM$?} If that was the case, it would account precisely for the additional two neutral and one charged scalars with electroweak-scale masses encountered in the numerical explorations.
The answer, no wonder, is positive. Out of inspired guessing, consider the complex vector
\begin{equation}\label{eq:NullVecCh:01}
 \vec c^{\,T}=
 \begin{pmatrix}
  \vev{1}e^{i2\vevPh{1}},& \ldots,& \vev{n}e^{i2\vevPh{n}}
 \end{pmatrix}.
\end{equation}
One can readily check that
\begin{equation}\label{eq:NullVecCh:02}
 \chMM\vec c=
 \begin{pmatrix}
  e^{i2\vevPh{1}}(\partial_{\vev{1}}{V}_2+\frac{i}{\vev{1}}\partial_{\vevPh{1}}{V}_2)\\
  \vdots\\
  e^{i2\vevPh{n}}(\partial_{\vev{n}}{V}_2+\frac{i}{\vev{n}}\partial_{\vevPh{n}}{V}_2)
 \end{pmatrix}.
\end{equation}
It is clear that $\vec c$ is indeed a null eigenvector of $\chMM$ owing to the stationarity conditions.
According to \refEq{eq:MChMN:Null:01}, $\vec r_1$ and $\vec r_2$ are null eigenvectors of $\nMM$, $\nMM\vec r_1=\nMM\vec r_2=\vec 0_{2n}$: 
\begin{equation}\label{eq:NullVecN:01}
\begin{aligned}
 &\vec r_1^{\,T}=(\re{\vec c}^T,-\im{\vec c}^T)=\\
 &(\vev{1}\cos 2\vevPh{1}, \ldots,\vev{n}\cos 2\vevPh{n}, -\vev{1}\sin 2\vevPh{1},  \ldots, -\vev{n}\sin 2\vevPh{n}),\\
 &\vec r_2^{\,T}=(\im{\vec c}^T,\re{\vec c}^T)=\\
 &(\vev{1}\sin 2\vevPh{1}, \ldots,\vev{n}\sin 2\vevPh{n}, \vev{1}\cos 2\vevPh{1},  \ldots, \vev{n}\cos 2\vevPh{n}).
\end{aligned}
\end{equation}
Although $\vec c$ is not orthogonal to $\vec c_G$, and correspondingly $\vec r_1$ and $\vec r_2$ are not orthogonal to $\vec r_G$ and $\vec r_h$ in \refEq{eq:NullVecNGold:01}, there is no worry on that respect: since they are independent, one can always orthonormalize \emph{\`a la} Gram-Schmidt.\\ 
Let us recap. Since masses much larger than the electroweak scale $\vev{}$ can only be obtained with large quadratic couplings $\gg\vev{}^2$ when quartic couplings are bounded by perturbativity constraints, we have analyzed the mass matrices in the absence of quartic couplings. We have found that in this regime, besides the expected null eigenvectors of $\chMM$ and $\nMM$ associated to the wbG and the Higgs-like scalar, there are, unexpectedly, further null eigenvectors, one of $\chMM$ and two of $\nMM$. One can now think of the complete picture, including quartic couplings as a perturbation with respect to the previous analysis --it is indeed degenerate perturbation theory that must be considered--. Besides the wbG which remain massless, it is thus clear that out of the remaining null eigenvectors of the ``no-quartics'' mass matrices, $\vec c$ in \refEq{eq:NullVecCh:01}, $\vec r_1$ and $\vec r_2$ in \refEq{eq:NullVecN:01}, together with $\vec r_h$ in \refEq{eq:NullVecNGold:01}, one charged and three neutral scalars get squared masses of order $(\QQ{}\text{'s})\times\vev{}^2$. They are \emph{light}: considering perturbativity requirements on the quartic couplings, their masses cannot be much larger than the electroweak scale, exactly as observed in our illustrative numerical exercise. Of course, more light states might be present when a regime other than ``all quadratic couplings are much larger than $\vev{}^2$'' is considered: the novel and relevant point is that with so few assumptions --a real $n$HDM with SCPV and bounded quartic couplings--, one can establish that, unexpectedly and no matter the number of doublets $n$, at least three new scalars \emph{must} be light.\\

One might reasonably wonder if the presence of the unexpected null eigenvectors is in some sense related to a broken continuous symmetry. Notice first that with the vacuum in \refEq{eq:Fieldvev:01}, for the CP conjugate vevs $\langle \SDc{a}\rangle  =\langle \SD{a}\rangle ^\ast $, we have $\potV(\langle \SD{1}\rangle ^\ast,\ldots,\langle \SD{n}\rangle ^\ast)=\potV(\langle \SD{1}\rangle ,\ldots,\langle \SD{n}\rangle)$, 
and thus $\langle \SD{a}\rangle ^\ast$ give a different candidate vacuum, obtained by CP-transforming the considered vacuum. When the quartic couplings are ignored, the mass terms in $\potV_2$ \emph{do not involve} vevs, and thus both the wbG corresponding to the vacuum and the wbG that would correspond to the CP transformed vacuum yield zero eigenvalues. 
Since no additional continuous symmetry appears to be spontaneously broken, the latter are not true wbG as these states are massive when the effects of $\potV_4$ are included. Their masses are bounded because from $\potV_2$ alone, they yield as good null mass eigenvectors as the true wbG.\footnote{The appearance of $e^{i2\vevPh{j}}$ phases in $\vec c$ in \refEq{eq:NullVecCh:01} compared to $\vec c_G$, without phases, is just an artifact due to the expansion of $\SD{a}$ in \refEq{eq:Fieldvev:01}, where the vacuum phase $e^{i\vevPh{a}}$ is explicitely factored out.}\\

One source of concern is if radiative corrections could alter the previous picture, potentially giving large corrections to the light masses. A detailed discussion on that respect is beyond the scope of this work, but one can nevertheless argue the following. Since the one loop effective scalar potential does not depend on the chosen vacuum and respects CP invariance, it is to be expected that our tree-level conclusions stand.\\

Considering the main result, phenomenological consequences --e.g. discovery potential at the LHC--, command attention. The question is beyond the scope of the present work for several reasons. Three types of interactions of these scalars should be considered: \emph{(i)} among scalars, arising from the quartic terms in \refEq{eq:RealnHDM:Pot:V4:01}, \emph{(ii)} Yukawa couplings to fermions, and \emph{(iii)} the ones arising from the covariant derivatives in the kinetic terms $(D_\mu\SD{a})^\dagger(D^\mu\SD{a})$. Considering the minimality of our assumptions and the large freedom available in both the interactions among the scalars and their couplings to fermions (which we have not addressed), the third type of interaction, involving scalars and gauge bosons, appears a priori better suited on that respect if observables insensitive to the first two types of interactions were available. There are, however, additional obstacles in that direction. Knowing that scalars with electroweak-scale masses (the eigenvalues of the mass matrices) might be present is not sufficient, a better understanding of the states (the corresponding eigenvectors) is necessary. Even in the regime where, apart from the necessarily light states, all other scalars are much heavier, these eigenvectors --which define the mixings in the scalar sector-- depend critically on the quartic couplings.
If one is not in that regime and there are more light states, the situation is even more involved, precluding at this stage a generic approach to guaranteed or clearly promising discovery prospects. 
It is to be stressed that in any case, nothing enforces that a priori one of the light states is precisely a SM-like Higgs boson, that is ``scalar alignment'' \cite{Gunion:2002zf,Carena:2013ooa,Pilaftsis:2016erj}. In general the would-be Higgs boson can have sizable mixing with the additional light scalars. Phenomenological input like for example $h\to WW^\ast$ signal strengths, puts significant constraints on that respect, imposing some tuning of parameters to force a SM-like Higgs boson in the spectrum.

\section{Conclusions\label{SEC:Conclusions}}
Extended scalar sectors, in particular multi-Higgs doublets models, featuring spontaneous electroweak symmetry breaking, necessarily include a Higgs-like state with mass not larger than the electroweak scale if perturbativity requirements are imposed on the potential. Owing to unconstrained quadratic couplings, one naive expectation is that, generically, all new scalars could be made arbitrarily heavy, with masses much larger than the electroweak scale. We have analyzed the case of real $n$HDM with SCPV, where contrary to such expectations and despite the abundance of free quadratic couplings, at least one charged and two additional neutral states have masses that cannot be larger than the electroweak scale, also due to perturbativity requirements on the quartic couplings in the potential.
This fact appears to be related to the existence of a different candidate vacuum, with both classes of minima of the potential related by a CP transformation: to which extent similar conclusions could apply to other scalar potentials is an open question.

\section*{Acknowledgements}
\textit{Conselleria de Innovación, Universidades, Ciencia y Sociedad Digital} from \textit{Generalitat Valenciana} (Spain) and \emph{Fondo Social Europeo} support MN and DQ through project CIDEGENT/2019/024, CM under grants ACIF/2021/284, CIBEFP/2022/92, and CIBEFP/2023/96. The authors are also supported by Spanish MICIU/AEI/10.13039/501100011033/ through grants PID2020-113334GB-I00, PID2023-151418NB-I00, and the \emph{Severo Ochoa} project CEX2023-001292-S. 


\begin{thebibliography}{30}%
\makeatletter
\providecommand \@ifxundefined [1]{%
 \@ifx{#1\undefined}
}%
\providecommand \@ifnum [1]{%
 \ifnum #1\expandafter \@firstoftwo
 \else \expandafter \@secondoftwo
 \fi
}%
\providecommand \@ifx [1]{%
 \ifx #1\expandafter \@firstoftwo
 \else \expandafter \@secondoftwo
 \fi
}%
\providecommand \natexlab [1]{#1}%
\providecommand \enquote  [1]{``#1''}%
\providecommand \bibnamefont  [1]{#1}%
\providecommand \bibfnamefont [1]{#1}%
\providecommand \citenamefont [1]{#1}%
\providecommand \href@noop [0]{\@secondoftwo}%
\providecommand \href [0]{\begingroup \@sanitize@url \@href}%
\providecommand \@href[1]{\@@startlink{#1}\@@href}%
\providecommand \@@href[1]{\endgroup#1\@@endlink}%
\providecommand \@sanitize@url [0]{\catcode `\\12\catcode `\$12\catcode
  `\&12\catcode `\#12\catcode `\^12\catcode `\_12\catcode `\%12\relax}%
\providecommand \@@startlink[1]{}%
\providecommand \@@endlink[0]{}%
\providecommand \url  [0]{\begingroup\@sanitize@url \@url }%
\providecommand \@url [1]{\endgroup\@href {#1}{\urlprefix }}%
\providecommand \urlprefix  [0]{URL }%
\providecommand \Eprint [0]{\href }%
\providecommand \doibase [0]{https://doi.org/}%
\providecommand \selectlanguage [0]{\@gobble}%
\providecommand \bibinfo  [0]{\@secondoftwo}%
\providecommand \bibfield  [0]{\@secondoftwo}%
\providecommand \translation [1]{[#1]}%
\providecommand \BibitemOpen [0]{}%
\providecommand \bibitemStop [0]{}%
\providecommand \bibitemNoStop [0]{.\EOS\space}%
\providecommand \EOS [0]{\spacefactor3000\relax}%
\providecommand \BibitemShut  [1]{\csname bibitem#1\endcsname}%
\let\auto@bib@innerbib\@empty
\bibitem [{\citenamefont {Aad}\ \emph {et~al.}(2012)\citenamefont {Aad} \emph
  {et~al.}}]{ATLAS:2012yve}%
  \BibitemOpen
  \bibfield  {author} {\bibinfo {author} {\bibfnamefont {G.}~\bibnamefont
  {Aad}} \emph {et~al.} (\bibinfo {collaboration} {ATLAS}),\ }\bibfield
  {title} {\bibinfo {title} {\emph{Observation of a new particle in the search for
  the Standard Model Higgs boson with the ATLAS detector at the LHC}},\ }\href
  {https://doi.org/10.1016/j.physletb.2012.08.020} {\bibfield  {journal}
  {\bibinfo  {journal} {Phys. Lett. B}\ }\textbf {\bibinfo {volume} {716}},\
  \bibinfo {pages} {1} (\bibinfo {year} {2012})},\ \Eprint
  {https://arxiv.org/abs/1207.7214} {arXiv:1207.7214 [hep-ex]} \BibitemShut
  {NoStop}%
\bibitem [{\citenamefont {Chatrchyan}\ \emph {et~al.}(2012)\citenamefont
  {Chatrchyan} \emph {et~al.}}]{CMS:2012qbp}%
  \BibitemOpen
  \bibfield  {author} {\bibinfo {author} {\bibfnamefont {S.}~\bibnamefont
  {Chatrchyan}} \emph {et~al.} (\bibinfo {collaboration} {CMS}),\ }\bibfield
  {title} {\bibinfo {title} {\emph{Observation of a New Boson at a Mass of 125 GeV
  with the CMS Experiment at the LHC}},\ }\href
  {https://doi.org/10.1016/j.physletb.2012.08.021} {\bibfield  {journal}
  {\bibinfo  {journal} {Phys. Lett. B}\ }\textbf {\bibinfo {volume} {716}},\
  \bibinfo {pages} {30} (\bibinfo {year} {2012})},\ \Eprint
  {https://arxiv.org/abs/1207.7235} {arXiv:1207.7235 [hep-ex]} \BibitemShut
  {NoStop}%
\bibitem [{\citenamefont {Baak}\ \emph {et~al.}(2012)\citenamefont {Baak},
  \citenamefont {Goebel}, \citenamefont {Haller}, \citenamefont {Hoecker},
  \citenamefont {Kennedy}, \citenamefont {Moenig}, \citenamefont {Schott},\
  and\ \citenamefont {Stelzer}}]{Baak:2011ze}%
  \BibitemOpen
  \bibfield  {author} {\bibinfo {author} {\bibfnamefont {M.}~\bibnamefont
  {Baak}}, \bibinfo {author} {\bibfnamefont {M.}~\bibnamefont {Goebel}},
  \bibinfo {author} {\bibfnamefont {J.}~\bibnamefont {Haller}}, \bibinfo
  {author} {\bibfnamefont {A.}~\bibnamefont {Hoecker}}, \bibinfo {author}
  {\bibfnamefont {D.}~\bibnamefont {Kennedy}}, \bibinfo {author} {\bibfnamefont
  {K.}~\bibnamefont {Moenig}}, \bibinfo {author} {\bibfnamefont
  {M.}~\bibnamefont {Schott}},\ and\ \bibinfo {author} {\bibfnamefont
  {J.}~\bibnamefont {Stelzer}} (\bibinfo {collaboration} {Gfitter}),\
  }\bibfield  {title} {\bibinfo {title} {\emph{Updated Status of the Global
  Electroweak Fit and Constraints on New Physics}},\ }\href
  {https://doi.org/10.1140/epjc/s10052-012-2003-4} {\bibfield  {journal}
  {\bibinfo  {journal} {Eur. Phys. J. C}\ }\textbf {\bibinfo {volume} {72}},\
  \bibinfo {pages} {2003} (\bibinfo {year} {2012})},\ \Eprint
  {https://arxiv.org/abs/1107.0975} {arXiv:1107.0975 [hep-ph]} \BibitemShut
  {NoStop}%
\bibitem [{\citenamefont {Weinberg}(1976)}]{Weinberg:1976pe}%
  \BibitemOpen
  \bibfield  {author} {\bibinfo {author} {\bibfnamefont {S.}~\bibnamefont
  {Weinberg}},\ }\bibfield  {title} {\bibinfo {title} {\emph{Mass of the Higgs
  Boson}},\ }\href {https://doi.org/10.1103/PhysRevLett.36.294} {\bibfield
  {journal} {\bibinfo  {journal} {Phys. Rev. Lett.}\ }\textbf {\bibinfo
  {volume} {36}},\ \bibinfo {pages} {294} (\bibinfo {year} {1976})}\BibitemShut
  {NoStop}%
\bibitem [{\citenamefont {Politzer}\ and\ \citenamefont
  {Wolfram}(1979)}]{Politzer:1978ic}%
  \BibitemOpen
  \bibfield  {author} {\bibinfo {author} {\bibfnamefont {H.~D.}\ \bibnamefont
  {Politzer}}\ and\ \bibinfo {author} {\bibfnamefont {S.}~\bibnamefont
  {Wolfram}},\ }\bibfield  {title} {\bibinfo {title} {\emph{Bounds on Particle
  Masses in the Weinberg-Salam Model}},\ }\href
  {https://doi.org/10.1016/0370-2693(79)90746-9} {\bibfield  {journal}
  {\bibinfo  {journal} {Phys. Lett. B}\ }\textbf {\bibinfo {volume} {82}},\
  \bibinfo {pages} {242} (\bibinfo {year} {1979})},\ \bibinfo {note} {[Erratum:
  Phys.Lett.B 83, 421 (1979)]}\BibitemShut {NoStop}%
\bibitem [{\citenamefont {Cabibbo}\ \emph {et~al.}(1979)\citenamefont
  {Cabibbo}, \citenamefont {Maiani}, \citenamefont {Parisi},\ and\
  \citenamefont {Petronzio}}]{Cabibbo:1979ay}%
  \BibitemOpen
  \bibfield  {author} {\bibinfo {author} {\bibfnamefont {N.}~\bibnamefont
  {Cabibbo}}, \bibinfo {author} {\bibfnamefont {L.}~\bibnamefont {Maiani}},
  \bibinfo {author} {\bibfnamefont {G.}~\bibnamefont {Parisi}},\ and\ \bibinfo
  {author} {\bibfnamefont {R.}~\bibnamefont {Petronzio}},\ }\bibfield  {title}
  {\bibinfo {title} {\emph{Bounds on the Fermions and Higgs Boson Masses in Grand
  Unified Theories}},\ }\href {https://doi.org/10.1016/0550-3213(79)90167-6}
  {\bibfield  {journal} {\bibinfo  {journal} {Nucl. Phys. B}\ }\textbf
  {\bibinfo {volume} {158}},\ \bibinfo {pages} {295} (\bibinfo {year}
  {1979})}\BibitemShut {NoStop}%
\bibitem [{\citenamefont {Dashen}\ and\ \citenamefont
  {Neuberger}(1983)}]{Dashen:1983ts}%
  \BibitemOpen
  \bibfield  {author} {\bibinfo {author} {\bibfnamefont {R.~F.}\ \bibnamefont
  {Dashen}}\ and\ \bibinfo {author} {\bibfnamefont {H.}~\bibnamefont
  {Neuberger}},\ }\bibfield  {title} {\bibinfo {title} {\emph{How to Get an Upper
  Bound on the Higgs Mass}},\ }\href
  {https://doi.org/10.1103/PhysRevLett.50.1897} {\bibfield  {journal} {\bibinfo
   {journal} {Phys. Rev. Lett.}\ }\textbf {\bibinfo {volume} {50}},\ \bibinfo
  {pages} {1897} (\bibinfo {year} {1983})}\BibitemShut {NoStop}%
\bibitem [{\citenamefont {Callaway}(1984)}]{Callaway:1983zd}%
  \BibitemOpen
  \bibfield  {author} {\bibinfo {author} {\bibfnamefont {D.~J.~E.}\
  \bibnamefont {Callaway}},\ }\bibfield  {title} {\bibinfo {title}
  {\emph{Nontriviality of Gauge Theories With Elementary Scalars and Upper Bounds on
  Higgs Masses}},\ }\href {https://doi.org/10.1016/0550-3213(84)90410-3}
  {\bibfield  {journal} {\bibinfo  {journal} {Nucl. Phys. B}\ }\textbf
  {\bibinfo {volume} {233}},\ \bibinfo {pages} {189} (\bibinfo {year}
  {1984})}\BibitemShut {NoStop}%
\bibitem [{\citenamefont {Lee}\ \emph {et~al.}(1977{\natexlab{a}})\citenamefont
  {Lee}, \citenamefont {Quigg},\ and\ \citenamefont {Thacker}}]{Lee:1977yc}%
  \BibitemOpen
  \bibfield  {author} {\bibinfo {author} {\bibfnamefont {B.~W.}\ \bibnamefont
  {Lee}}, \bibinfo {author} {\bibfnamefont {C.}~\bibnamefont {Quigg}},\ and\
  \bibinfo {author} {\bibfnamefont {H.~B.}\ \bibnamefont {Thacker}},\
  }\bibfield  {title} {\bibinfo {title} {\emph{The Strength of Weak Interactions at
  Very High-Energies and the Higgs Boson Mass}},\ }\href
  {https://doi.org/10.1103/PhysRevLett.38.883} {\bibfield  {journal} {\bibinfo
  {journal} {Phys. Rev. Lett.}\ }\textbf {\bibinfo {volume} {38}},\ \bibinfo
  {pages} {883} (\bibinfo {year} {1977}{\natexlab{a}})}\BibitemShut {NoStop}%
\bibitem [{\citenamefont {Lee}\ \emph {et~al.}(1977{\natexlab{b}})\citenamefont
  {Lee}, \citenamefont {Quigg},\ and\ \citenamefont {Thacker}}]{Lee:1977eg}%
  \BibitemOpen
  \bibfield  {author} {\bibinfo {author} {\bibfnamefont {B.~W.}\ \bibnamefont
  {Lee}}, \bibinfo {author} {\bibfnamefont {C.}~\bibnamefont {Quigg}},\ and\
  \bibinfo {author} {\bibfnamefont {H.~B.}\ \bibnamefont {Thacker}},\
  }\bibfield  {title} {\bibinfo {title} {\emph{Weak Interactions at Very
  High-Energies: The Role of the Higgs Boson Mass}},\ }\href
  {https://doi.org/10.1103/PhysRevD.16.1519} {\bibfield  {journal} {\bibinfo
  {journal} {Phys. Rev. D}\ }\textbf {\bibinfo {volume} {16}},\ \bibinfo
  {pages} {1519} (\bibinfo {year} {1977}{\natexlab{b}})}\BibitemShut {NoStop}%
\bibitem [{\citenamefont {Dicus}\ and\ \citenamefont
  {Mathur}(1973)}]{Dicus:1973gbw}%
  \BibitemOpen
  \bibfield  {author} {\bibinfo {author} {\bibfnamefont {D.~A.}\ \bibnamefont
  {Dicus}}\ and\ \bibinfo {author} {\bibfnamefont {V.~S.}\ \bibnamefont
  {Mathur}},\ }\bibfield  {title} {\bibinfo {title} {\emph{Upper bounds on the
  values of masses in unified gauge theories}},\ }\href
  {https://doi.org/10.1103/PhysRevD.7.3111} {\bibfield  {journal} {\bibinfo
  {journal} {Phys. Rev. D}\ }\textbf {\bibinfo {volume} {7}},\ \bibinfo {pages}
  {3111} (\bibinfo {year} {1973})}\BibitemShut {NoStop}%
\bibitem [{\citenamefont {Langacker}\ and\ \citenamefont
  {Weldon}(1984)}]{Langacker:1984dma}%
  \BibitemOpen
  \bibfield  {author} {\bibinfo {author} {\bibfnamefont {P.}~\bibnamefont
  {Langacker}}\ and\ \bibinfo {author} {\bibfnamefont {H.~A.}\ \bibnamefont
  {Weldon}},\ }\bibfield  {title} {\bibinfo {title} {\emph{A Mass Sum Rule for Higgs
  Bosons in Arbitrary Models}},\ }\href
  {https://doi.org/10.1103/PhysRevLett.52.1377} {\bibfield  {journal} {\bibinfo
   {journal} {Phys. Rev. Lett.}\ }\textbf {\bibinfo {volume} {52}},\ \bibinfo
  {pages} {1377} (\bibinfo {year} {1984})}\BibitemShut {NoStop}%
\bibitem [{\citenamefont {Weldon}(1984)}]{Weldon:1984th}%
  \BibitemOpen
  \bibfield  {author} {\bibinfo {author} {\bibfnamefont {H.~A.}\ \bibnamefont
  {Weldon}},\ }\bibfield  {title} {\bibinfo {title} {\emph{Constraints on Scalar
  Masses Implied by Spontaneous Symmetry Breaking}},\ }\href
  {https://doi.org/10.1016/0370-2693(84)90643-9} {\bibfield  {journal}
  {\bibinfo  {journal} {Phys. Lett. B}\ }\textbf {\bibinfo {volume} {146}},\
  \bibinfo {pages} {59} (\bibinfo {year} {1984})}\BibitemShut {NoStop}%
\bibitem [{\citenamefont {Lee}(1973)}]{Lee:1973iz}%
  \BibitemOpen
  \bibfield  {author} {\bibinfo {author} {\bibfnamefont {T.~D.}\ \bibnamefont
  {Lee}},\ }\bibfield  {title} {\bibinfo {title} {\emph{A Theory of Spontaneous T
  Violation}},\ }\href {https://doi.org/10.1103/PhysRevD.8.1226} {\bibfield
  {journal} {\bibinfo  {journal} {Phys. Rev. D}\ }\textbf {\bibinfo {volume}
  {8}},\ \bibinfo {pages} {1226} (\bibinfo {year} {1973})}\BibitemShut
  {NoStop}%
\bibitem [{\citenamefont {Lee}(1974)}]{Lee:1974jb}%
  \BibitemOpen
  \bibfield  {author} {\bibinfo {author} {\bibfnamefont {T.~D.}\ \bibnamefont
  {Lee}},\ }\bibfield  {title} {\bibinfo {title} {\emph{CP Nonconservation and
  Spontaneous Symmetry Breaking}},\ }\href
  {https://doi.org/10.1016/0370-1573(74)90020-9} {\bibfield  {journal}
  {\bibinfo  {journal} {Phys. Rept.}\ }\textbf {\bibinfo {volume} {9}},\
  \bibinfo {pages} {143} (\bibinfo {year} {1974})}\BibitemShut {NoStop}%
\bibitem [{\citenamefont {Huffel}\ and\ \citenamefont
  {Pocsik}(1981)}]{Huffel:1980sk}%
  \BibitemOpen
  \bibfield  {author} {\bibinfo {author} {\bibfnamefont {H.}~\bibnamefont
  {Huffel}}\ and\ \bibinfo {author} {\bibfnamefont {G.}~\bibnamefont
  {Pocsik}},\ }\bibfield  {title} {\bibinfo {title} {\emph{Unitarity Bounds on Higgs
  Boson Masses in the Weinberg-Salam Model With Two Higgs Doublets}},\ }\href
  {https://doi.org/10.1007/BF01429824} {\bibfield  {journal} {\bibinfo
  {journal} {Z. Phys. C}\ }\textbf {\bibinfo {volume} {8}},\ \bibinfo {pages}
  {13} (\bibinfo {year} {1981})}\BibitemShut {NoStop}%
\bibitem [{\citenamefont {Casalbuoni}\ \emph {et~al.}(1988)\citenamefont
  {Casalbuoni}, \citenamefont {Dominici}, \citenamefont {Feruglio},\ and\
  \citenamefont {Gatto}}]{Casalbuoni:1987cz}%
  \BibitemOpen
  \bibfield  {author} {\bibinfo {author} {\bibfnamefont {R.}~\bibnamefont
  {Casalbuoni}}, \bibinfo {author} {\bibfnamefont {D.}~\bibnamefont
  {Dominici}}, \bibinfo {author} {\bibfnamefont {F.}~\bibnamefont {Feruglio}},\
  and\ \bibinfo {author} {\bibfnamefont {R.}~\bibnamefont {Gatto}},\ }\bibfield
   {title} {\bibinfo {title} {\emph{Tree Level Unitarity Violation for Large Scalar
  Mass in Multi-Higgs Extensions of the Standard Model}},\ }\href
  {https://doi.org/10.1016/0550-3213(88)90469-5} {\bibfield  {journal}
  {\bibinfo  {journal} {Nucl. Phys. B}\ }\textbf {\bibinfo {volume} {299}},\
  \bibinfo {pages} {117} (\bibinfo {year} {1988})}\BibitemShut {NoStop}%
\bibitem [{\citenamefont {Maalampi}\ \emph {et~al.}(1991)\citenamefont
  {Maalampi}, \citenamefont {Sirkka},\ and\ \citenamefont
  {Vilja}}]{Maalampi:1991fb}%
  \BibitemOpen
  \bibfield  {author} {\bibinfo {author} {\bibfnamefont {J.}~\bibnamefont
  {Maalampi}}, \bibinfo {author} {\bibfnamefont {J.}~\bibnamefont {Sirkka}},\
  and\ \bibinfo {author} {\bibfnamefont {I.}~\bibnamefont {Vilja}},\ }\bibfield
   {title} {\bibinfo {title} {\emph{Tree level unitarity and triviality bounds for
  two Higgs models}},\ }\href {https://doi.org/10.1016/0370-2693(91)90068-2}
  {\bibfield  {journal} {\bibinfo  {journal} {Phys. Lett. B}\ }\textbf
  {\bibinfo {volume} {265}},\ \bibinfo {pages} {371} (\bibinfo {year}
  {1991})}\BibitemShut {NoStop}%
\bibitem [{\citenamefont {Kanemura}\ \emph {et~al.}(1993)\citenamefont
  {Kanemura}, \citenamefont {Kubota},\ and\ \citenamefont
  {Takasugi}}]{Kanemura:1993hm}%
  \BibitemOpen
  \bibfield  {author} {\bibinfo {author} {\bibfnamefont {S.}~\bibnamefont
  {Kanemura}}, \bibinfo {author} {\bibfnamefont {T.}~\bibnamefont {Kubota}},\
  and\ \bibinfo {author} {\bibfnamefont {E.}~\bibnamefont {Takasugi}},\
  }\bibfield  {title} {\bibinfo {title} {\emph{Lee-Quigg-Thacker bounds for Higgs
  boson masses in a two doublet model}},\ }\href
  {https://doi.org/10.1016/0370-2693(93)91205-2} {\bibfield  {journal}
  {\bibinfo  {journal} {Phys. Lett. B}\ }\textbf {\bibinfo {volume} {313}},\
  \bibinfo {pages} {155} (\bibinfo {year} {1993})},\ \Eprint
  {https://arxiv.org/abs/hep-ph/9303263} {arXiv:hep-ph/9303263} \BibitemShut
  {NoStop}%
\bibitem [{\citenamefont {Ginzburg}\ and\ \citenamefont
  {Ivanov}(2005)}]{Ginzburg:2005dt}%
  \BibitemOpen
  \bibfield  {author} {\bibinfo {author} {\bibfnamefont {I.~F.}\ \bibnamefont
  {Ginzburg}}\ and\ \bibinfo {author} {\bibfnamefont {I.~P.}\ \bibnamefont
  {Ivanov}},\ }\bibfield  {title} {\bibinfo {title} {\emph{Tree-level unitarity
  constraints in the most general 2HDM}},\ }\href
  {https://doi.org/10.1103/PhysRevD.72.115010} {\bibfield  {journal} {\bibinfo
  {journal} {Phys. Rev. D}\ }\textbf {\bibinfo {volume} {72}},\ \bibinfo
  {pages} {115010} (\bibinfo {year} {2005})},\ \Eprint
  {https://arxiv.org/abs/hep-ph/0508020} {arXiv:hep-ph/0508020} \BibitemShut
  {NoStop}%
\bibitem [{\citenamefont {Horejsi}\ and\ \citenamefont
  {Kladiva}(2006)}]{Horejsi:2005da}%
  \BibitemOpen
  \bibfield  {author} {\bibinfo {author} {\bibfnamefont {J.}~\bibnamefont
  {Horejsi}}\ and\ \bibinfo {author} {\bibfnamefont {M.}~\bibnamefont
  {Kladiva}},\ }\bibfield  {title} {\bibinfo {title} {\emph{Tree-unitarity bounds
  for THDM Higgs masses revisited}},\ }\href
  {https://doi.org/10.1140/epjc/s2006-02472-3} {\bibfield  {journal} {\bibinfo
  {journal} {Eur. Phys. J. C}\ }\textbf {\bibinfo {volume} {46}},\ \bibinfo
  {pages} {81} (\bibinfo {year} {2006})},\ \Eprint
  {https://arxiv.org/abs/hep-ph/0510154} {arXiv:hep-ph/0510154} \BibitemShut
  {NoStop}%
\bibitem [{\citenamefont {Kanemura}\ and\ \citenamefont
  {Yagyu}(2015)}]{Kanemura:2015ska}%
  \BibitemOpen
  \bibfield  {author} {\bibinfo {author} {\bibfnamefont {S.}~\bibnamefont
  {Kanemura}}\ and\ \bibinfo {author} {\bibfnamefont {K.}~\bibnamefont
  {Yagyu}},\ }\bibfield  {title} {\bibinfo {title} {\emph{Unitarity bound in the
  most general two Higgs doublet model}},\ }\href
  {https://doi.org/10.1016/j.physletb.2015.10.047} {\bibfield  {journal}
  {\bibinfo  {journal} {Phys. Lett. B}\ }\textbf {\bibinfo {volume} {751}},\
  \bibinfo {pages} {289} (\bibinfo {year} {2015})},\ \Eprint
  {https://arxiv.org/abs/1509.06060} {arXiv:1509.06060 [hep-ph]} \BibitemShut
  {NoStop}%
\bibitem [{\citenamefont {Ivanov}(2017)}]{Ivanov:2017dad}%
  \BibitemOpen
  \bibfield  {author} {\bibinfo {author} {\bibfnamefont {I.~P.}\ \bibnamefont
  {Ivanov}},\ }\bibfield  {title} {\bibinfo {title} {\emph{Building and testing
  models with extended Higgs sectors}},\ }\href
  {https://doi.org/10.1016/j.ppnp.2017.03.001} {\bibfield  {journal} {\bibinfo
  {journal} {Prog. Part. Nucl. Phys.}\ }\textbf {\bibinfo {volume} {95}},\
  \bibinfo {pages} {160} (\bibinfo {year} {2017})},\ \Eprint
  {https://arxiv.org/abs/1702.03776} {arXiv:1702.03776 [hep-ph]} \BibitemShut
  {NoStop}%
\bibitem [{\citenamefont {Haber}\ and\ \citenamefont
  {Nir}(1990)}]{Haber:1989xc}%
  \BibitemOpen
  \bibfield  {author} {\bibinfo {author} {\bibfnamefont {H.~E.}\ \bibnamefont
  {Haber}}\ and\ \bibinfo {author} {\bibfnamefont {Y.}~\bibnamefont {Nir}},\
  }\bibfield  {title} {\bibinfo {title} {\emph{Multiscalar Models With a High-energy
  Scale}},\ }\href {https://doi.org/10.1016/0550-3213(90)90499-4} {\bibfield
  {journal} {\bibinfo  {journal} {Nucl. Phys. B}\ }\textbf {\bibinfo {volume}
  {335}},\ \bibinfo {pages} {363} (\bibinfo {year} {1990})}\BibitemShut
  {NoStop}%
\bibitem [{\citenamefont {Faro}\ \emph {et~al.}(2020)\citenamefont {Faro},
  \citenamefont {Romao},\ and\ \citenamefont {Silva}}]{Faro:2020qyp}%
  \BibitemOpen
  \bibfield  {author} {\bibinfo {author} {\bibfnamefont {F.}~\bibnamefont
  {Faro}}, \bibinfo {author} {\bibfnamefont {J.~C.}\ \bibnamefont {Romao}},\
  and\ \bibinfo {author} {\bibfnamefont {J.~P.}\ \bibnamefont {Silva}},\
  }\bibfield  {title} {\bibinfo {title} {\emph{Nondecoupling in Multi-Higgs doublet
  models}},\ }\href {https://doi.org/10.1140/epjc/s10052-020-8217-y} {\bibfield
   {journal} {\bibinfo  {journal} {Eur. Phys. J. C}\ }\textbf {\bibinfo
  {volume} {80}},\ \bibinfo {pages} {635} (\bibinfo {year} {2020})},\ \Eprint
  {https://arxiv.org/abs/2002.10518} {arXiv:2002.10518 [hep-ph]} \BibitemShut
  {NoStop}%
\bibitem [{\citenamefont {Carrolo}\ \emph {et~al.}(2021)\citenamefont
  {Carrolo}, \citenamefont {Rom\~ao}, \citenamefont {Silva},\ and\
  \citenamefont {Vaz\~ao}}]{Carrolo:2021euy}%
  \BibitemOpen
  \bibfield  {author} {\bibinfo {author} {\bibfnamefont {S.}~\bibnamefont
  {Carrolo}}, \bibinfo {author} {\bibfnamefont {J.~C.}\ \bibnamefont
  {Rom\~ao}}, \bibinfo {author} {\bibfnamefont {J.~a.~P.}\ \bibnamefont
  {Silva}},\ and\ \bibinfo {author} {\bibfnamefont {F.}~\bibnamefont
  {Vaz\~ao}},\ }\bibfield  {title} {\bibinfo {title} {\emph{Symmetry and decoupling
  in multi-Higgs boson models}},\ }\href
  {https://doi.org/10.1103/PhysRevD.103.075026} {\bibfield  {journal} {\bibinfo
   {journal} {Phys. Rev. D}\ }\textbf {\bibinfo {volume} {103}},\ \bibinfo
  {pages} {075026} (\bibinfo {year} {2021})},\ \Eprint
  {https://arxiv.org/abs/2102.11303} {arXiv:2102.11303 [hep-ph]} \BibitemShut
  {NoStop}%
\bibitem [{\citenamefont {Nebot}\ \emph {et~al.}(2019)\citenamefont {Nebot},
  \citenamefont {Botella},\ and\ \citenamefont {Branco}}]{Nebot:2018nqn}%
  \BibitemOpen
  \bibfield  {author} {\bibinfo {author} {\bibfnamefont {M.}~\bibnamefont
  {Nebot}}, \bibinfo {author} {\bibfnamefont {F.~J.}\ \bibnamefont {Botella}},\
  and\ \bibinfo {author} {\bibfnamefont {G.~C.}\ \bibnamefont {Branco}},\
  }\bibfield  {title} {\bibinfo {title} {\emph{Vacuum Induced CP Violation
  Generating a Complex CKM Matrix with Controlled Scalar FCNC}},\ }\href
  {https://doi.org/10.1140/epjc/s10052-019-7221-6} {\bibfield  {journal}
  {\bibinfo  {journal} {Eur. Phys. J. C}\ }\textbf {\bibinfo {volume} {79}},\
  \bibinfo {pages} {711} (\bibinfo {year} {2019})},\ \Eprint
  {https://arxiv.org/abs/1808.00493} {arXiv:1808.00493 [hep-ph]} \BibitemShut
  {NoStop}%
\bibitem [{\citenamefont {Nebot}(2020)}]{Nebot:2019qvr}%
  \BibitemOpen
  \bibfield  {author} {\bibinfo {author} {\bibfnamefont {M.}~\bibnamefont
  {Nebot}},\ }\bibfield  {title} {\bibinfo {title} {\emph{Bounded masses in two
  Higgs doublets models, spontaneous CP violation and $Z_2$ symmetry}},\ }\href
  {https://doi.org/10.1103/PhysRevD.102.115002} {\bibfield  {journal} {\bibinfo
   {journal} {Phys. Rev. D}\ }\textbf {\bibinfo {volume} {102}},\ \bibinfo
  {pages} {115002} (\bibinfo {year} {2020})},\ \Eprint
  {https://arxiv.org/abs/1911.02266} {arXiv:1911.02266 [hep-ph]} \BibitemShut
  {NoStop}%
\bibitem [{\citenamefont {Nierste}\ \emph {et~al.}(2020)\citenamefont
  {Nierste}, \citenamefont {Tabet},\ and\ \citenamefont
  {Ziegler}}]{Nierste:2019fbx}%
  \BibitemOpen
  \bibfield  {author} {\bibinfo {author} {\bibfnamefont {U.}~\bibnamefont
  {Nierste}}, \bibinfo {author} {\bibfnamefont {M.}~\bibnamefont {Tabet}},\
  and\ \bibinfo {author} {\bibfnamefont {R.}~\bibnamefont {Ziegler}},\
  }\bibfield  {title} {\bibinfo {title} {\emph{Cornering Spontaneous CP Violation
  with Charged-Higgs-Boson Searches}},\ }\href
  {https://doi.org/10.1103/PhysRevLett.125.031801} {\bibfield  {journal}
  {\bibinfo  {journal} {Phys. Rev. Lett.}\ }\textbf {\bibinfo {volume} {125}},\
  \bibinfo {pages} {031801} (\bibinfo {year} {2020})},\ \Eprint
  {https://arxiv.org/abs/1912.11501} {arXiv:1912.11501 [hep-ph]} \BibitemShut
  {NoStop}%
\bibitem [{\citenamefont {Barenboim}\ \emph {et~al.}(2002)\citenamefont
  {Barenboim}, \citenamefont {Gorbahn}, \citenamefont {Nierste},\ and\
  \citenamefont {Raidal}}]{Barenboim:2001vu}%
  \BibitemOpen
  \bibfield  {author} {\bibinfo {author} {\bibfnamefont {G.}~\bibnamefont
  {Barenboim}}, \bibinfo {author} {\bibfnamefont {M.}~\bibnamefont {Gorbahn}},
  \bibinfo {author} {\bibfnamefont {U.}~\bibnamefont {Nierste}},\ and\ \bibinfo
  {author} {\bibfnamefont {M.}~\bibnamefont {Raidal}},\ }\bibfield  {title}
  {\bibinfo {title} {\emph{Higgs Sector of the Minimal Left-Right Symmetric
  Model}},\ }\href {https://doi.org/10.1103/PhysRevD.65.095003} {\bibfield
  {journal} {\bibinfo  {journal} {Phys. Rev. D}\ }\textbf {\bibinfo {volume}
  {65}},\ \bibinfo {pages} {095003} (\bibinfo {year} {2002})},\ \Eprint
  {https://arxiv.org/abs/hep-ph/0107121} {arXiv:hep-ph/0107121} \BibitemShut
  {NoStop}%
\bibitem [{\citenamefont {Gunion}\ and\ \citenamefont
  {Haber}(2003)}]{Gunion:2002zf}%
  \BibitemOpen
  \bibfield  {author} {\bibinfo {author} {\bibfnamefont {J.~F.}\ \bibnamefont
  {Gunion}}\ and\ \bibinfo {author} {\bibfnamefont {H.~E.}\ \bibnamefont
  {Haber}},\ }
  \bibfield  {title}
  {\bibinfo {title} {\emph{The CP conserving two Higgs doublet model: The Approach to the decoupling limit}},\ }
  \href {https://doi.org/10.1103/PhysRevD.67.075019} {\bibfield
  {journal} {\bibinfo  {journal} {Phys. Rev. D}\ }\textbf {\bibinfo {volume}
  {67}},\ \bibinfo {pages} {075019} (\bibinfo {year} {2003})},\ \Eprint
  {https://arxiv.org/abs/hep-ph/0207010} {arXiv:hep-ph/0207010} \BibitemShut
  {NoStop}%
\bibitem [{\citenamefont {Carena}\ \emph {et~al.}(2014)\citenamefont {Carena},
  \citenamefont {Low}, \citenamefont {Shah},\ and\ \citenamefont
  {Wagner}}]{Carena:2013ooa}%
  \BibitemOpen
  \bibfield  {author} {\bibinfo {author} {\bibfnamefont {M.}~\bibnamefont
  {Carena}}, \bibinfo {author} {\bibfnamefont {I.}~\bibnamefont {Low}},
  \bibinfo {author} {\bibfnamefont {N.~R.}\ \bibnamefont {Shah}},\ and\
  \bibinfo {author} {\bibfnamefont {C.~E.~M.}\ \bibnamefont {Wagner}},\ }
  \bibfield  {title}
  {\bibinfo {title} {\emph{Impersonating the Standard Model Higgs Boson: Alignment without Decoupling}},\ }
  \href
  {https://doi.org/10.1007/JHEP04(2014)015} {\bibfield  {journal} {\bibinfo
  {journal} {JHEP}\ }\textbf {\bibinfo {volume} {04}},\ \bibinfo {pages}
  {015}},\ \Eprint {https://arxiv.org/abs/1310.2248} {arXiv:1310.2248 [hep-ph]}
  \BibitemShut {NoStop}%
\bibitem [{\citenamefont {Pilaftsis}(2016)}]{Pilaftsis:2016erj}%
  \BibitemOpen
  \bibfield  {author} {\bibinfo {author} {\bibfnamefont {A.}~\bibnamefont
  {Pilaftsis}},\ }
    \bibfield  {title}
  {\bibinfo {title} {\emph{Symmetries for standard model alignment in multi-Higgs doublet models}},\ }
  \href {https://doi.org/10.1103/PhysRevD.93.075012} {\bibfield
   {journal} {\bibinfo  {journal} {Phys. Rev. D}\ }\textbf {\bibinfo {volume}
  {93}},\ \bibinfo {pages} {075012} (\bibinfo {year} {2016})},\ \Eprint
  {https://arxiv.org/abs/1602.02017} {arXiv:1602.02017 [hep-ph]} \BibitemShut
  {NoStop}%
\end{thebibliography}
%

\end{document}